\documentclass[11pt,1p]{elsarticle}
\usepackage{dsfont, amsmath, verbatim, graphicx}
\usepackage{epsfig}
\usepackage{bbm}
\usepackage[latin1]{inputenc} 

\biboptions{sort&compress} 

\newcommand{\ket}[1]{|#1\rangle}
\newcommand{\ketbra}[1]{| #1\rangle \langle #1|}

\newcommand{\be}{\begin{equation}}
\newcommand{\ee}{\end{equation}}
\newcommand{\eea}{\end{eqnarray}}
\newcommand{\bea}{\begin{eqnarray}}

\newcommand{\eins}{\mathbbm{1}}

\newcommand{\HH}{\ensuremath{\mathcal{H}}}
\newcommand{\FF}{\ensuremath{\mathcal{F}}}

\renewcommand{\AA}{\ensuremath{\mathcal{A}}}
\newcommand{\BB}{\ensuremath{\mathcal{B}}}

\newcommand{\LL}{\ensuremath{\mathcal{L}}}

\newcommand{\kommentar}[1]{}

\newcommand{\vr}{\ensuremath{\varrho}}

\newcommand{\forget}[1]{}

\begin{document}
\title{\LARGE Entanglement criteria and full separability of multi-qubit  quantum states}
\author{Otfried G\"uhne}
\address{Institut f\"ur Quantenoptik and Quanteninformation, \"Osterreichische Akademie der Wissenschaften, Technikerstr.~21A, 6020 Innsbruck, Austria}
\address{Institut f\"ur Theoretische Physik, Universit\"at Innsbruck, Technikerstr.~25, 6020 Innsbruck, Austria}
\address{Fachbereich Physik, Universit\"at Siegen, Walter-Flex-Str.~3, 57068 Siegen, Germany }

\begin{abstract}
We introduce an entanglement criterion to exclude full separability of quantum states. 
We present numerical evidence that the criterion is necessary and sufficient 
for the class of GHZ diagonal three-qubit states and estimate the volume of 
bound entangled states within this class. Finally, we extend our approach to 
bound entangled states which are not GHZ diagonal.
\end{abstract}

\maketitle

\section{Introduction}

Entanglement is believed to be an important resource in quantum information 
processing and consequently many works are devoted to its characterization 
\cite{hororeview,gtreview}. This characterization becomes complicated, if more 
than two particles are entangled, since then different classes of entanglement 
exist. Some methods to distinguish between the different classes have been
presented in the literature \cite{duer99,horoalt,doherty,yusong,wcjan,clarisse, 
hassanyoag,zukowski, tothss, gs09,huber,niekamp,gittsovich,kay10,kay10neu}. However, no general 
solution of the problem is known, not even for specific families of states.

In this paper we present a criterion for the verification of multi-qubit
entanglement. The criterion is formulated as a set of simple inequalities 
for the matrix elements of the state. Our approach is inspired by some existing 
criteria \cite{gs09},
and also by some recent works on entanglement in the family of graph states 
\cite{kay10,kay10neu}. We consider then density matrices which are diagonal 
in terms of Greenberger-Horne-Zeilinger (GHZ) states and investigate the 
optimality of
our criterion. For that, we derive methods to prove that a given state is 
separable and it turns out that all GHZ diagonal states under scrutiny are
either detected by our criterion or proven to be separable. This allows to 
estimate the volume of so-called bound entangled states in this class of 
states.
Bound entanglement is a weak form of entanglement, where no pure state 
entanglement can be distilled from; and this phenomenon is central to 
many open problems in quantum information theory \cite{hororeview}.
Finally, we discuss with the help of an example how our ideas can be 
used to characterize bound entanglement close to the three-qubit W state.

\section{Definitions and statement of the problem}
Before explaining our separability criteria, we introduce 
the notation and give some examples of existing 
separability criteria and interesting quantum states.

\subsection{Separability and entanglement}
We consider an $N$-particle system with Hilbert space 
$\HH_{\rm tot}=\HH^{\otimes N}.$ 
Any matrix $\vr$ acting on $\HH_{\rm tot}$ which is hermitian ($\vr=\vr^\dagger$), has no 
negative eigenvalues ($\vr\geq 0$) and is normalized [$Tr(\vr)=1$] is a valid 
density matrix of some quantum state. By definition, a state is fully separable, 
if it can be written as a convex combination of product states,
\be
\vr = \sum_k p_k \ketbra{a_k} \otimes \ketbra{b_k} \otimes ... \otimes \ketbra{z_k},
\label{sepdef}
\ee
where the $p_k$ are non-negative ($p_k\geq 0$) and normalized  ($\sum_k p_k = 1$); 
in other words, they form a probability distribution. If a state cannot be written 
as in Eq.~(\ref{sepdef}) it is entangled in some sense. In general, it is not easy to
check whether a given quantum state is fully separable or not, see 
Refs.~\cite{duer99,horoalt, doherty, yusong, wcjan, clarisse, hassanyoag, zukowski, tothss, gs09, huber} for 
some existing sufficient criteria for entanglement and Refs.~\cite{plenio1,barreiro} for numerical 
tests to prove separability. Especially if a state is only weakly entangled, proving 
entanglement is not straightforward.

In this paper we will derive criteria, which allow to prove that a state is not fully
separable and hence contains some entanglement. It should be stressed that there are
more refined notions of entanglement for multiparticle systems (e.g., genuine multipartite 
entanglement) which can be more relevant for special situations (e.g., experiments). 
We will, however, concentrate on full separability as the basic definition, for a 
discussion of recent results on the other classifications see the reviews in 
Refs.~\cite{hororeview, gtreview}. Moreover, we will focus our discussion on the case of 
three qubits, but our results can directly be generalized to more particles.

\subsection{A criterion for full separability in terms of the matrix elements}
Let us explain a separability criterion for full separability derived in Ref.~\cite{gs09}, 
which we will generalize later. For three qubits, consider the $8\times 8$-density matrix 
with entries $\vr_{i,j}.$ Here and in the following, we always order the basis vectors in 
the canonical way, $\{\ket{000},\ket{001}, \ket{010},..., \ket{111}\}.$
Then, if the state is fully separable, the entries fulfill
\be
|\vr_{1,8}| \leq \sqrt[6]{\vr_{2,2}\vr_{3,3}\vr_{4,4} \vr_{5,5}\vr_{6,6}\vr_{7,7}}.
\label{oldcriterion}
\ee
The idea of the proof is as follows: It is easy to check that for a pure fully 
separable state equality holds in Eq.~(\ref{oldcriterion}). Then, $|\vr_{1,8}|$ is convex in 
$\vr$ [a function $f(\vr)$ is convex if $f[p\vr_1+(1-p)\vr_2] \leq p f(\vr_1)+(1-p)f(\vr_2)$],
while the right-hand side of Eq.~(\ref{oldcriterion}) is concave 
[i.e., $f[p\vr_1+(1-p)\vr_2] \geq p f(\vr_1)+(1-p)f(\vr_2)$]. This implies 
the bound for mixed fully separable states.

The right-hand side of the inequality (\ref{oldcriterion}) may be replaced by other 
expressions, for instance, $|\vr_{1,8}| \leq \sqrt[4]{\vr_{2,2}\vr_{3,3} \vr_{5,5}\vr_{8,8}}$
holds for separable states. Then, this criterion is able to detect the entanglement in a 
family of states which are separable with respect to any bipartition, but not fully 
separable \cite{gs09}. This is then a bound entangled state and the criterion improves existing 
criteria for this family significantly.

In the following, we will extend the criterion in Eq.~(\ref{oldcriterion}) by considering
more than one off-diagonal element on the left-hand side. We will also discuss the optimal
choice of the right-hand side.

\subsection{A bound entangled state}
Let us now introduce a bound entangled state which motivates our approach. 
Consider the family of three-qubit density matrices introduced by 
A.~Kay \cite{kay10},
\be
\vr_{\rm AK}(\hat \alpha) =
\frac{1}{8+8\hat \alpha}
\begin{pmatrix}
4 +\hat \alpha & 0 &0  & 0 & 0 &0 & 0 & 2
\\
0 & \hat \alpha &  0 & 0 & 0 & 0 &  2 & 0  
\\
0 & 0 &  \hat \alpha & 0 & 0 & -2 &  0 & 0 
\\
0 & 0 &  0 & \hat \alpha & 2 & 0 &  0 & 0  
\\
0 & 0 &  0 & 2 & \hat \alpha & 0 &  0 & 0  
\\
0 & 0 &  -2 & 0 & 0 & \hat \alpha &  0 & 0  
\\
0 & 2 &  0 & 0 & 0 & 0 &  \hat \alpha & 0  
\\
2 & 0 &  0 & 0 & 0 & 0 &  0 &  4 +\hat \alpha
\end{pmatrix}.
\label{kaystate}
\ee
This matrix is a valid quantum state for $\hat \alpha \geq 2$
and it is  separable for $\hat \alpha \geq 2 \sqrt{2}.$ This has been
proved in Ref.~\cite{kay10} by writing down an 
explicit separable decomposition as in Eq.~(\ref{sepdef}).
Furthermore, using the algorithm 
outlined in Ref.~\cite{doherty} it was shown numerically that 
the state is entangled for $2 \leq \hat \alpha \leq 2.828.$ Note 
that for any $\hat \alpha \geq 2$ the state has a positive partial 
transpose (PPT) for any bipartition. This does not only imply that no 
entanglement can be distilled from it\footnote{At this point, one 
should mention that undistillable entanglement can arise in 
multiparticle systems in a simple way: If $\vr$ is entangled 
with respect to one bipartition, but PPT with respect to another 
bipartition, this implies already that $\vr$ is entangled, but 
multipartite undistillable. In this paper, however, we use a more 
restrictive definition: we consider bound entangled states, that 
are separable for any bipartition, but not fully separable. This 
means that they are undistillable, even if arbitrary parties join.};
moreover, since the state is diagonal in the GHZ basis (see also below), 
it also implies that the state is separable for all bipartitions. Namely, from 
the fact that a two-qubit Bell diagonal state with a positive 
partial transpose is separable, one can conclude that a GHZ 
diagonal state that has a positive partial transposition is
also separable  for that partition \cite{nagata}.

The state in Eq.~(\ref{kaystate}) is an example of a GHZ diagonal state.
These are of the form 
\be
\vr= \sum_{k=1}^8 p_k \ketbra{GHZ_k},
\label{ghzdia1}
\ee
where the GHZ state basis consists of the eight vectors 
$\ket{GHZ_k} = \ket{0 x_2 x_3} \pm \ket{1 \bar{x}_2 \bar{x}_3}$ where $x_i, \bar{x}_i \in \{0,1\}$ 
and $x_i \neq \bar{x}_i.$ Alternatively, the GHZ diagonal
states can be written as
\be
\vr= \frac{1}{8}
\big[
\eins + \lambda_2 ZZ\eins + \lambda_3 Z\eins Z + \lambda_4 \eins Z Z +
\lambda_5 XXX + \lambda_6 YYX + \lambda_7 YXY  + \lambda_8 XYY  
\big],
\label{ghzdia2}
\ee
where $X,Y,$ and $Z$ denote the the Pauli matrices and tensor product 
symbols have been omitted. The observables occurring in Eq.~(\ref{ghzdia2})
are the so-called stabilizing observables of the GHZ states, see 
Ref.~\cite{hein} for a further discussion.

GHZ diagonal states have been intensively discussed in the literature before 
\cite{duer99, gs09, nagata, pittenger} and are interesting from several perspectives: They 
have a simple structure, since only the diagonal and the anti-diagonal elements 
of the matrix can be nonzero, and  they occur naturally in certain types of decoherence 
processes \cite{aolita}. Furthermore, any state can be transformed to a GHZ diagonal 
state (without changing the fidelities of the GHZ states) by local operations 
\cite{duer99, hein}. This means that if the remaining GHZ diagonal state is entangled, the 
initial state must have been entangled, too. On the other hand, if one has a 
GHZ diagonal state $\vr_1$ and a second separable state $\vr_2$ which is mapped
to $\vr_1$ by these local operations, then $\vr_1$ must be separable, too.

The state $\vr_{\rm AK}$ is not detected by the criterion in Eq.~(\ref{oldcriterion})
or variants thereof. The reason lies in the fact that Eq.~(\ref{oldcriterion}) 
considers only single offdiagonal elements, and neglects the (phase) relations 
between them. It is the main purpose of this paper to develop an improvement of 
Eq.~(\ref{oldcriterion}) which takes into account all offdiagonal elements at the 
same time. This will finally prove analytically that the state $\vr_{\rm AK}$ is 
entangled if $2 \leq \hat \alpha < 2 \sqrt{2}.$

\section{The separability criterion}
In this section we will now formulate the separability criterion. We will 
restrict our attention to three qubits, but it should be stressed that 
our approach can straightforwardly be generalized to an arbitrary number 
of qubits. 

To start, consider a pure product state, 
\be
\ket{\phi} = (c_1 \ket{0} + s_1 \ket{1}) \otimes (c_2 \ket{0} + s_2 \ket{1})
\otimes (c_3 \ket{0} + s_3 \ket{1}).
\label{pps}
\ee
Here, the complex coefficients $c_i$ and $s_i$ fulfill the normalization 
$|c_i|^2+|s_i|^2=1.$ Let us consider the corresponding $8 \times 8$ density 
matrix $\vr=\ketbra{\phi}.$ The offdiagonal elements can be written as
\bea
&&\vr_{1,8} = c_1 c_2 c_3 s_1^* s_2^* s_3^*  = \kappa  e^{i(a+b+c)},
\nonumber
\\
&&\vr_{2,7} = c_1 c_2 s_3 s_1^* s_2^* c_3^*  = \kappa  e^{i a},
\nonumber
\\
&&\vr_{3,6} = c_1 s_2 c_3 s_1^* c_2^* s_3^*  = \kappa  e^{i b},
\nonumber
\\
&&\vr_{5,4} = s_1 c_2 c_3 c_1^* s_2^* s_3^*  = \kappa  e^{i c}.
\label{offdiagonals}
\eea
Here, we used the notations $a= \phi_1 + \phi_2 - \phi_3,$
$b= \phi_1 - \phi_2 + \phi_3,$ and 
$c= - \phi_1 + \phi_2 + \phi_3,$ where the phases are defined via
$c_k s_k^*= |c_k s_k| e^{i \phi_k}.$ Furthermore, we set
$\kappa=|c_1 s_1||c_2 s_2||c_3 s_3|.$ Note that the other offdiagonal 
elements follow from these and the fact that $\vr$ is hermitian.

From Eq.~(\ref{offdiagonals}) one can conclude two things. First, 
for a product states the absolute value of all the offdiagonal elements 
is the same. Second, the four phases are not arbitrary; they depend 
only on three parameters.

To proceed, consider a linear functional like 
\be
\LL(\vr,\vec{X}) = \Re(X_1 \vr_{1,8} + X_2 \vr_{2,7} + X_3 \vr_{3,6} +X_4 \vr_{5,4}),
\ee
where $\vec{X}=(X_1,X_2,X_3,X_4)$ is a vector of complex coefficients and 
$\Re(...)$ denotes the real part. The functional $\LL$ is compatible with 
convex combinations of the quantum state, i.e., one has 
$\LL[p\vr_1 + (1-p)\vr_2]=p\LL(\vr_1)+(1-p)\LL(\vr_2).$
For a pure separable state, $\LL$ is given by
\bea
\LL(\ket{\phi})&=&\kappa \FF(\vec{X}) \mbox{ with }
\nonumber
\\
\FF(\vec{X}) &=&
\Re(X_1)\cos(a+b+c)-\Im(X_1)\sin(a+b+c) + 
\Re(X_2)\cos(a)
\nonumber
\\
&&-\Im(X_2)\sin(a) + 
\Re(X_3)\cos(b)-\Im(X_3)\sin(b) 
\nonumber
\\&&+ 
\Re(X_4)\cos(c)-\Im(X_4)\sin(c).
\eea
In order to obtain an extension of the separability condition in 
Eq.~(\ref{oldcriterion}) we need two more facts. First, for given 
coefficients $\vec{X}$ one may compute the maximum given by
\be
C(\vec{X})=\sup_{a,b,c} |\FF(\vec{X})| 
\label{fdef}
\ee
For given values of $\vec{X}$ this can usually be computed analytically (see also below)
or with a simple numerical optimization.

Second, we have to characterize $\kappa$ for product states. For them we have
\bea
\kappa &=&\sqrt[4]{\vr_{1,1}\vr_{4,4} \vr_{6,6}\vr_{7,7}}
=
\sqrt[4]{\vr_{2,2}\vr_{3,3} \vr_{5,5}\vr_{8,8}}
\nonumber
\\
&=&
\sqrt{\vr_{k,k}\vr_{9-k,9-k}} \mbox{ for all $1\leq k\leq 4$}.
\label{kappas}
\eea
There are further equalities of this type (e.g.,
$\kappa=\sqrt[6]{\vr_{2,2}\vr_{3,3}\vr_{4,4} \vr_{5,5}\vr_{6,6}\vr_{7,7}}$)
and in view of Eq.~(\ref{oldcriterion}) one might be tempted to use them.
This, however, will not give stronger criteria: $\kappa$ will be used to 
deliver an upper bound, and in general a bound like 
$x \leq \sqrt[3]{\alpha \beta \gamma}$ [corresponding to Eq.~(\ref{oldcriterion})]
is weaker than the bound $x \leq \min\{\alpha,\beta,\gamma\}$
[corresponding to the last line in Eq.~(\ref{kappas})]. In a similar way 
one can directly see that other possible bounds (as used in Ref.~\cite{gs09})
can be derived from the terms in Eq.~(\ref{kappas}).

Putting it all together, we can formulate:

{\bf Observation.}
{\it Let $\vec{X}$ be some coefficients and $C(\vec{X})$ be defined as in Eq.~(\ref{fdef}).
Then, if $\vr$ is fully separable the inequality
\bea
|\LL(\vr,\vec{X})| &\leq& C(\vec{X}) 
\min
\big\{\sqrt[4]{\vr_{1,1}\vr_{4,4} \vr_{6,6}\vr_{7,7}},
\sqrt[4]{\vr_{2,2}\vr_{3,3} \vr_{5,5}\vr_{8,8}},
\nonumber
\\
&&
\sqrt{\vr_{k,k}\vr_{9-k,9-k}}\;\;(\mbox{with } 1 \leq k \leq 4)
\big\}
\label{newcriterion}
\eea
holds and violation of this inequality implies entanglement.}

To prove this criterion, note first that Eq.~(\ref{newcriterion}) holds 
for pure product states. Furthermore, the left-hand side is a convex function 
of the state, while 
the right-hand side is concave
\cite{gs09}. Since a mixed fully separable state is a convex combination of fully separable
pure states, Eq.~(\ref{newcriterion}) has to hold.

It remains to discuss which parameters $\vec{X}$ should be chosen in order
to detect a given state $\vr.$ A simple choice is
$\vec{X} \sim (\vr_{1,8}^*,\vr_{2,7}^*,\vr_{3,6}^*,\vr_{5,4}^*),$ since then 
the Cauchy-Schwarz inequality guarantees that $\LL(\vr,\vec{X})$ is maximal 
for all $\vec{X}$ with the same normalization. The optimal choice, however, is
to choose $\vec{X}$ such that $|\LL(\vr,\vec{X})|/C(\vec{X})$ is maximal.

\section{Examples}

\subsection{The bound entangled state from Eq.~(\ref{kaystate})}
As a first example, let us discuss the state $\vr_{\rm AK}$ from 
Eq.~(\ref{kaystate}). First, let us take a look at
\be
C[\vec{X}=(\delta,\alpha,\beta,\gamma)]=\sup_{a,b,c}[\delta\cos(a+b+c)+\alpha\cos(a)+\beta\cos(b)+\gamma\cos(c)],
\ee
where the coefficients $\delta,\alpha,\beta,\gamma$ are real. This case is important for GHZ-diagonal states, since for them all anti-diagonal elements are real.

If these coefficients are all positive, it is clear that the maximum is $C= |\delta| +|\alpha|+|\beta|+ |\gamma|$ and is attained
at $a=b=c=0.$ The same value for $C$ is obtained, if two or four of the coefficients are negative, 
because with transformations like $a\mapsto a+\pi$ one can flip two arbitrary signs of the cosine 
terms. 

If one of the coefficients is negative (we choose $\gamma <0$ for definiteness) 
one finds after some algebra that the optimum can be attained at
\bea
\sin(a) = \frac{1}{2 \alpha}\frac{\sqrt{Q}}{\sqrt{R}}, && \cos(a)=\pm \sqrt{1-\sin(a)^2},
\nonumber
\\
\sin(b) = \frac{1}{2 \beta}\frac{\sqrt{Q}}{\sqrt{R}}, && \cos(b)=\pm \sqrt{1-\sin(b)^2},
\nonumber
\\
\sin(c)=\frac{1}{2 \gamma}\frac{\sqrt{Q}}{\sqrt{R}}, && \cos(c)=\pm \sqrt{1-\sin(c)^2},
\eea
with
\bea
Q&=&-
(\alpha \beta \delta + \alpha \beta \gamma - \alpha \delta \gamma - 
          \beta \delta \gamma)
(\alpha \beta \delta - \alpha \beta \gamma + 
        \alpha \delta \gamma - \beta \delta \gamma)
\nonumber
\\
\nonumber
&&
(\alpha \beta \delta - 
        \alpha \beta \gamma - \alpha \delta \gamma + 
        \beta \delta \gamma)(\alpha \beta \delta + \alpha \beta \gamma + 
        \alpha \delta \gamma + \beta \delta \gamma),
\\
R &=& \alpha \beta \gamma \delta 
(\alpha \beta -  \delta \gamma)
(\alpha \gamma - \beta \delta) 
(\alpha \delta - \beta \gamma).
\eea
This is the optimal solution, if the $\sin(...)$ are real and their absolute value
is not larger than one. Furthermore, one has to distribute maximally one negative 
sign in the $\cos(...)$ terms such that the condition $\delta \sin(a+b+c)+\alpha\sin(a)=0$ 
(originating from setting the derivative to zero) holds. If the conditions
on $\sin(...)$ are not met, the optimum is just given by a choice of $\{a,b,c\}\in \{0,\pi\}$
resulting in $C=\alpha+\beta+\gamma+\delta$ or $C=\alpha+\beta-\gamma-\delta$ etc. The solution
is of the same structure, if three of the coefficients $\alpha,\beta,\gamma,\delta$ are negative.

Applying this to the state from Eq.~(\ref{kaystate}) one finds that
$
C[\vec{X}=(1,1,-1,1)]=2\sqrt{2}.
$
According to Eq.~(\ref{newcriterion}), the state is entangled, 
if $8 > 2 \sqrt{2} \hat\alpha \Leftrightarrow \hat\alpha < 2 \sqrt{2},$
which proves the numerical result. Recall that for larger values of
$\hat\alpha$ the state can be proven to be separable (see Ref.~\cite{kay10} 
and also below), so the criterion is optimal.

\subsection{Random GHZ diagonal states}
As a second example, we consider randomly chosen GHZ diagonal states [see also 
Eqs.~(\ref{ghzdia1}, \ref{ghzdia2})]. Here, we also want to estimate how good the 
criteria are, in the sense that we want to test whether some entangled states escape
from the detection.\footnote{For results on certain subclasses of GHZ diagonal states
see Refs.~\cite{kay10, kay10neu, pittenger}.}

Consequently, we need a method to prove that a given state is separable. This 
is, in general, not straightforward, but there are two possibilities that we 
used:
First, in Ref.~\cite{barreiro} a simple iterative algorithm was proposed 
that can be used to prove different forms of separability for general 
quantum states.

Second, for the special case of GHZ diagonal states one can also use the ideas 
proposed recently in Refs.~\cite{kay10,kay10neu}. Let us explain them step by 
step. First, a state like $\vr \sim \alpha \eins + \alpha Z Z \eins $ is 
separable, as it is diagonal in a product basis and has no negative eigenvalues.
Then, consider an operator $\AA=\lambda_2 ZZ\eins + \lambda_3 Z\eins Z + \lambda_4 \eins Z Z.$
The minimal eigenvalue of it is given by 
$\lambda_-
=\min\{\lambda_2+\lambda_3+\lambda_4,\lambda_2-\lambda_3-\lambda_4,-\lambda_2+\lambda_3-\lambda_4,
-\lambda_2-\lambda_3+\lambda_4\}$ and $\lambda_-$ is never positive. Therefore, 
a state like $\vr \sim |\lambda_-| \eins + \AA$ is also separable, as it is 
diagonal in the computational basis and positive semidefinite.

Considering the state in Eq.~(\ref{ghzdia2}), this allows already to conclude that a GHZ diagonal 
state is separable, if $|\lambda_-|+|\lambda_5|+|\lambda_6|+|\lambda_7|+|\lambda_8| \leq 1.$ However, 
inspired by Ref.~\cite{kay10neu} we can go further: taking the operator
$\BB=\lambda_5 XXX + \lambda_6 YYX + \lambda_7 YXY  + \lambda_8 XYY$ we may 
try to rewrite it as
\be
\BB = \frac{1}{2}
\big[
(A_1+B_1)\otimes(A_2+B_2)\otimes(A_3+B_3)
+(A_1-B_1)\otimes(A_2-B_2)\otimes(A_3-B_3)
\big],
\label{zerlegung}
\ee
with $A_i= \mu \cos(\vartheta_i)X$ and $B_i= \mu \sin(\vartheta_i)Y$. Then, the minimal 
eigenvalue of each 
of the terms in the right-hand side of Eq.~(\ref{zerlegung}) is given by $-\mu^3,$ and 
the matrix $\vr \sim \mu^3\eins+\BB$ is separable.

If all $\lambda_i$ in $\BB$ are positive, a solution of Eq.~(\ref{zerlegung}) can be 
found with
\be
\mu^3(\lambda_5, \lambda_6, \lambda_7, \lambda_8)=
\frac{
\sqrt{
(\lambda_5 \lambda_6 + \lambda_7 \lambda_8)
(\lambda_5 \lambda_7+\lambda_6 \lambda_8)
(\lambda_5 \lambda_8+ \lambda_6 \lambda_7)}}
{\sqrt{\lambda_5 \lambda_6 \lambda_7 \lambda_8}}
\label{muloesung}
\ee
and the same solution holds if an even number of $\lambda_i$ are negative. Therefore, 
in this  case the  state is separable, if 
$|\lambda_-|+ \mu^3 \leq 1.$ 
For 
the case that an odd number of the $\lambda_i$ in $\BB$ is negative, a 
decomposition as in Eq.~(\ref{zerlegung}) will lead to non-hermitian $A_i$ and 
$B_i$ which does not help. Then, only the separability condition
$|\lambda_-|+|\lambda_5|+|\lambda_6|+|\lambda_7|+|\lambda_8| \leq 1$
remains. But the case that $\prod_{i=5}^{8}\lambda_i < 0$ is exactly the 
case, for which it has been shown already in Ref.~\cite{kay10} that the PPT
criterion is necessary and sufficient for full separability (note the different
sign conventions in Ref.~\cite{kay10}).

In some cases, these conditions can still be improved. For that, one 
may consider operators of the form
\be
\mathcal{X} = 
p (A_1+B_1)\otimes(A_2+B_2)\otimes(A_3+B_3)
+(1-p)(\hat{A}_1-\hat B_1)\otimes(\hat A_2-\hat B_2)\otimes(\hat A_3-\hat B_3)
\label{besserezerlegung}
\ee
with $A_i=\alpha_i X$, $B_i=\beta_i Y$, $\hat A_i=\hat \alpha_i X$, $\hat B_i=\hat \beta_i Y$,
and $0 \leq p \leq 1.$ The minimal eigenvalues of the two terms are 
$\eta=-\prod_i(\sqrt{\alpha_i^2+\beta_i^2})$ and $\hat \eta=-\prod_i(\sqrt{\hat \alpha_i^2+\hat \beta_i^2})$,
respectively. Therefore, the state 
\be
\vr_{\rm sep} \sim (p|\eta|+(1-p)|\hat\eta|)\eins + \mathcal{X}
\label{besserezerlegung2}
\ee 
is separable.

This state $\vr_{\rm sep}$ is not diagonal in 
the GHZ basis as it contains also terms like $XXY$ etc. But, by applying local operations 
one can make it GHZ diagonal without changing the weights of $XXX,XYY,YXY,$ and $YYX.$
In fact, after making the terms in Eq.~\ref{besserezerlegung} GHZ diagonal, a convex 
combination of two times Eq.~\ref{zerlegung} arises.

Consequently, 
if we consider a given GHZ diagonal state $\vr_{\rm ghz}=\chi \eins + \BB$ and find a separable state 
like $\vr_{\rm sep}$ with the same weights 
[that is, $ p\alpha_1\alpha_2 \alpha_3 + (1-p)\hat \alpha_1\hat \alpha_2 \hat \alpha_3 =\lambda_5$ 
etc.] then $\vr_{\rm ghz}$ must be separable if $\chi \geq [p|\eta|+(1-p)|\hat\eta|].$ The search for the appropriate
$\mathcal{X}$ can easily be done numerically. This criterion is stronger than the one of Eqs.~(\ref{zerlegung}, \ref{muloesung})
since it contains the latter as the special case $p=1/2.$

To test our criteria, we have generated $10^6$ GHZ diagonal states, by choosing the 
eigenvalues $p_k$ in Eq.~(\ref{ghzdia1}) 
randomly from the seven-dimensional simplex in $\mathbbm{R}^8$. 
Then, we tested the criterion of 
the positivity of the partial transposition \cite{hororeview, gtreview} as well as our new criterion 
from Eq.~(\ref{newcriterion}). For this criterion we found the optimal $\vec{X}$ via a simple numerical
optimization. Note that for GHZ diagonal states the criterion 
in Eq.~(\ref{newcriterion}) is strictly stronger than the PPT criterion: For instance, 
a state that is PPT with respect to the  $A|BC$ partition has to fulfill 
$|\vr_{18}|\leq \sqrt{\vr_{44}\vr_{55}}$ which is a special case of Eq.~(\ref{newcriterion}).
For the states that were not detected by these criteria, we have tried to prove that they are separable
using the ideas from above. The results are given in Table I.

\begin{table}
\begin{center}
\begin{tabular}{|l|r||l|r|}
\hline
Entangled  &  91.32 \% &
NPT for some partition  & 90.61 \%  
\\
\cline{3-4}
{[via Eq.~(\ref{newcriterion})]} &  &
PPT, but violating Eq.~(\ref{newcriterion})& 0.71 \%  \\
\hline
Separable & 8.68 \% &
Via Eqs.~(\ref{zerlegung}, \ref{muloesung})& 8.41 \%
\\
\cline{3-4}
&&
Using in addition
Eqs.~(\ref{besserezerlegung}, \ref{besserezerlegung2})& 0.27 \%
\\
\hline
\hline
Total & 100 \% &  & 100 \%
\\
\hline
\end{tabular}
\end{center}
\caption{Fractions of the randomly generated states, which are detected by the different
criteria. See the text for details.}
\end{table}
First, it is important that {\it any} state which was not detected by the new criterion 
in Eq.~(\ref{newcriterion}) was proven to be separable. This gives clear evidence
that the criterion is a necessary and sufficient entanglement criterion for GHZ 
diagonal states. Moreover, the volume of the bound entangled states among the GHZ
diagonal states can be estimated to be $0.7\%.$ 

Finally, we also compared the general separability algorithm from Ref.~\cite{barreiro} 
with the special criteria for GHZ diagonal states in Eqs.~(\ref{zerlegung}--\ref{besserezerlegung2}). 
It turns out that the general algorithm succeeds for $96\%$ of the separable states and is therefore 
a powerful tool to prove separability, which can also be used if a state is not GHZ diagonal.

\section{Bound entanglement in the vicinity of the W state}
The previous criterion was well suited for GHZ diagonal states. For 
other states, one may first apply local transformations in order to 
bring them close to a GHZ diagonal state. 
Nevertheless, this will not always succeed and one can therefore ask 
whether similar ideas can be used for other bound entangled states, 
which are not GHZ diagonal. To see that this is possible, let us 
consider the state investigated by P. Hyllus \cite{hyllusdiss},
\be
\vr_{\rm PH}(\eta) =
\frac{1}{3+2\eta +3/ \eta}
\begin{pmatrix}
2\eta & 0 &0  & 0 & 0 &0 & 0 & 0
\\
0 & 1 &  1 & 0 & 1 & 0 &  0 & 0  
\\
0 & 1 &  1 & 0 & 1 & 0 &  0 & 0 
\\
0 & 0 &  0 & 1/\eta & 0 & 0 &  0 & 0  
\\
0 & 1 &  1 & 0 & 1 & 0 &  0 & 0  
\\
0 & 0 &  0 & 0 & 0 & 1/\eta &  0 & 0  
\\
0 & 0 &  0 & 0 & 0 & 0 &  1/\eta & 0  
\\
0 & 0 &  0 & 0 & 0 & 0 &  0 & 0
\end{pmatrix}.
\label{hyllusstate}
\ee
This state is close to the three-qubit W state 
$\ket{W}=(\ket{001}+ \ket{010} + \ket{100})/\sqrt{3},$
and is separable with respect to any bipartition, but not 
fully separable. The entanglement in this state is not detected by
the methods presented in Ref.~\cite{gs09}.

To investigate the entanglement properties of these states, consider 
first filter operations of the form 
\be
\FF_{\rm tot} = \FF \otimes \FF \otimes \FF \mbox{ with } 
\FF=
\begin{pmatrix} 
\tfrac{1}{{x}} &0 
\\
0 & x^2
\end{pmatrix}.
\ee
Under this filtering operation the state transforms like 
$\vr_{\rm PH}(\eta)\mapsto \FF_{\rm tot}[\vr_{\rm PH}(\eta)]\FF_{\rm tot} $ $\sim \vr_{\rm PH}(\eta/x^6).$
Therefore, all states in this family share the same entanglement properties and one may focus on the
case that $\eta=\sqrt{3/2}$, where the off-diagonal terms are maximal, 
$|\vr_{2,3}|+|\vr_{3,5}|+|\vr_{5,2}|=1/(1+\sqrt{8/3})\approx 0.38$.

In order to derive a separability criterion, we consider a pure product state as in 
Eq.~(\ref{pps}) with $|s_1|^2|s_2|^2|s_3|^2 = \vr_{8,8} \leq \varepsilon.$ Then, for 
some $i \in \{1,2,3\}$ the bound $|s_i|\leq \sqrt[6]{\varepsilon}$ must hold, and from that it follows after 
a short calculation (using $|c_j|\leq 1$) that 
$|\vr_{2,3}|+|\vr_{3,5}|+|\vr_{5,2}| \leq \sqrt[6]{\varepsilon}+1/4.$
Therefore, any pure separable state obeys
\be
|\vr_{2,3}|+|\vr_{3,5}|+|\vr_{5,2}| \leq \sqrt[6]{\vr_{8,8}}+1/4.
\ee
The left-hand side of this inequality is convex, while the right-hand side is concave, so this
inequality holds also for mixed separable states. The state $\vr_{\rm PH}$ with $\eta=\sqrt{3/2}$
clearly violates it (also if some noise is added), so the states in the family $\vr_{\rm PH}(\eta)$ 
must be entangled.

\section{Conclusion}
In conclusion, we have presented a criterion for full separability 
in terms of inequalities for the density matrix elements. The criterion 
seems to be necessary and sufficient for GHZ diagonal three-qubit states 
and allows to characterize bound entangled states. Finally, we showed that 
our ideas can also be extended to different families of bound entangled states
close to the W state. 

There are several possibilities to extend our work in the future. First, 
an analytical proof of the necessity of the criterion for GHZ diagonal states would be desireable. 
Then, an explicit investigation of the criteria for four or more qubits would be of 
interest. Finally, one could try to extend our analysis to the more general class of 
states which are diagonal in a graph state basis.

We thank B. Jungnitsch, P. Hyllus, A. Kay, M. Kleinmann, T. Moroder, S. Niekamp 
and M. Seevinck for discussions. This work has been supported 
by the FWF (START Prize and SFB FOQUS) and the EU (NAMEQUAM).

\end{document}